\documentstyle[preprint,revtex]{aps}
\tightenlines
\begin{document}
\draft

\preprint{UTF 300}

\begin{title}
Free and self-interacting scalar fields\\
in the presence of conical singularities
\end{title}
\author{Guido Cognola\cite{cognola}}
\begin{instit}
Dipartimento di Fisica, Universit\`a di Trento \\
and
Istituto Nazionale di Fisica Nucleare \\
Gruppo Collegato di Trento, Italia
\end{instit}
\author{Klaus Kirsten\cite{kirsten}}
\begin{instit}
Dipartimento di Fisica, Universit\`a di Trento, Italia
\end{instit}
\author{Luciano Vanzo\cite{vanzo}}
\begin{instit}
Dipartimento di Fisica, Universit\`a di Trento \\
and
Istituto Nazionale di Fisica Nucleare \\
Gruppo Collegato di Trento, Italia
\end{instit}

\date{\today}

\begin{abstract}
Free and self-interacting scalar fields in the presence of
conical singularities are analized in some detail.
The role of such a kind of singularities on free and vacuum
energy and also on the one-loop effective action is pointed out
using $\zeta$-function regularization and heat-kernel techniques.
\end{abstract}

\section{Introduction}
Quantum field theory in curved spacetime has been an important and
interesting tool as a first step towards quantum gravity during the
lets say last two decades. Concentrating first on free scalar and
higher spin fields (the main results may be found in
Birrell and Davies \cite{birelldavies82}),
afterwards also much progress has been made in
self-interacting theories on curved spacetime
(for a general review see \cite{buchbinderodintsovshapiro92}).
For example the
renormalizability of self-interacting $\phi^4$ scalar field theory in
four dimensional curved spacetime has been shown
(see for example
\cite{nelsonpanangaden82,toms82}).
Furthermore,
in connection with inflationary cosmological models the relevance
of the gravitational background field for the effective potential has
been considered in detail
\cite{oconnorhu84,critchleyhustylianopoulos87,buchbinderodintsovshapiro92}.

In all these considerations the so called heat-kernel coefficients
play a central role because they determine the pole structure of the
Green functions of the considered theory and thus the necessary
counterterms in the corresponding effective Lagrangians (at least to
one-loop)  \cite{wiesendangerwipfunv}.

Recent progress on heat-kernel coefficients allowed the generalization
of the mentioned considerations to curved spacetimes with boundary. In
\cite{bransongilkey90} (see also
\cite{dettkiwipf92,cognolavanzozerbini90,mcavityosborn91,mossdowker89,dowkerschofield90})
the magic $a_2$-coefficient has been
determined for a second order elliptic operator, the leading symbol
of which
is the metric tensor for a general smooth manifold with smooth
boundary. As an application the Casimir energy on this kind of
manifolds has been analysed for free fields
\cite{elizalderomeo90,kirsten91b}
and the (one-loop) renormalization program has been performed
in a self-interacting scalar field theory \cite{odintsov90a,odintsov91}.

Nowadays, quantum field theory on non-smooth manifolds like orbifolds
is of increasing interest
\cite{changdowker92,changdowker93,fursaevmieleunv,aurellsalomonsonunv}.
In general, it is very difficult to
obtain the heat-kernel coefficients or the $\zeta$-functions of the
relevant operators in such a kind of manifolds. Therefore, because at
the moment there is no general scheme available, we concentrate in
this article on the influence of the simplest non-smoothness a
manifold may have, namely a conical singularity. It is realized by
introducing a periodic coordinate ranging from $0$ to $\beta$, which
can be a real angle (as in a physical wedge) or imaginary time (as in
the Rindler wedge). Some important work on quantum field theory of a
free massless particle has already been done
\cite{dowker77,dowker78}. We will
generalize this work to the massive scalar field and to
self-interacting scalar field theory on a cone. The influence of the
conical singularity in the different contexts will be considered in
this paper.

More detailed the organization of the paper is as follows.
In Sec.~\ref{S:ZFR} we briefly summarize the $\zeta$-function regularization
technique as introduced by Hawking in Ref.~\cite{hawking77} and use it
in Sec.~\ref{S:VE} in order to obtain the contribution
to the vacuum energy due to the conical singularity, in the case of a
massive scalar field, and in Sec.~\ref{S:FT}, in order to obtain finite
temperature effects.
In Sec.~\ref{S:RW} we focus our attention on the interesting case
of the Rindler wedge, in which the conical singularity is due
to the appearance of a horizon.
In Sec.~\ref{S:RLP4} we describe in detail how the presence of a
singular point modify the one loop effective action for a
self-interacting scalar field and finally, in Sec.~\ref{S:C},
we conclude with some comments and possible applications of the result
we have obtained.
The spectral geometry of the cone is given in the two
Appendices, in which we also describe all
mathematical apparatus we need.

Throughout the paper we will use units in which $\hbar=c=1$.

\section{$\zeta$-function regularization}
\label{S:ZFR}

In this section we will briefly introduce the $\zeta$-function
regularization scheme
\cite{hawking77,critchleydowker76}
used in order to define the relevant physical
quantities. The considered spacetimes ${\cal M}$ will be
$N$-dimensional ultrastatic ones, thus
we present the basic equations needed in sections
\ref{S:VE}, ~\ref{S:FT} and \ref{S:RW}
for a finite temperature quantum field
theory of a free scalar field propagating in this kind of spacetimes.
In Sec.~\ref{S:RLP4} we will have to add some part due to the
self-interacting potential.

Finite temperature is incorporated using the Euclidean time formalism.
Thus we perform a Wick rotation of the time variable $x^0$ to the
imaginary time $\tau =ix^0$, requiring the scalar field $\phi$ to be
periodic in imaginary time with period $\beta$, where $\beta =\frac 1
T$ is the inverse temperature.

The relevant second order elliptic differential operator describing
the propagation of a free
scalar field is given by $L_N=-\triangle_N+m^2+\xi R$
together with appropriate boundary conditions for the field,
where $\triangle_N$ is
the Laplace-Beltrami operator on the manifold ${\cal M}$ and $m$ is
the mass of the field. In terms of $L_N$ the partition function at
temperature $T$ in the $\zeta$-function scheme is defined to be
\begin{eqnarray}
Z_\beta=\int_{\phi(0,\vec x)=\phi(\beta,\vec x)}d[\phi]\;
\exp\left(-\frac12\int_{{\cal M}}\phi L_N\phi d^Nx\right)
=\exp(\frac{1}{2}\zeta'_\beta(0|L_N/\mu^2))
\label{PF}
\end{eqnarray}
where $\mu$ is an arbitrary renormalization
parameter coming from the path integral measure.
All regularized physical quantities then can be directly derived from
Eq.~(\ref{PF}) using the usual formulae of thermodynamics.
As usual we call $Z_\beta$ partition function, but in fact it differs
from the thermodynamical one for the presence of the vacuum energy.
The corresponding zero-temperature results are
obtained in the limit $\beta\to\infty$.

Let us mention that
in the cases we shall consider in the paper, the manifold ${\cal M}$
will be the direct product between a $(N-2)$ dimensional manifold
and a cone ${\cal C}_\gamma$ with semi-angle
$\sin^{-1}(\gamma/2\pi)$ with coordinates $\bar r\equiv(r,\varphi)$.
In this way, the field operator separates in
$L_N=-\triangle_{N-2}+\triangle_c+m^2+\xi R$, $\triangle_c$ being the Laplacian
on ${\cal C}_\gamma$ (see Eq.~(\ref{33})).

For ultrastatic manifolds one obtains
$L_N=-\partial_\tau^2+L_{N-1}$ and after some algebraic manipulation
using the Jacoby identity \cite{hille62}
one has \cite{cognolavanzozerbini92}
\begin{eqnarray}
\zeta_\beta(s|L_N/\mu^2)
&=&\frac{\mu\beta\Gamma(s-1/2)}{\sqrt{4\pi}\Gamma(s)}
\zeta(s-1/2|L_{N-1}/\mu^2)\nonumber\\
&&+\frac{\mu\beta}{\sqrt{\pi}\Gamma(s)}
\sum_{n=1}^{\infty}\int_{0}^{\infty}t^{s-3/2}e^{-(n\mu\beta)^2/4t}
\,\mbox{Tr}\, e^{-tL_{N-1}/\mu^2}\;dt
\label{}
\end{eqnarray}
from which, easily follows
\begin{eqnarray}
\ln Z_\beta&=&-\frac{\beta}{2}\left[\bar\zeta(-1/2|L_{N-1})
-\frac{\bar\mu K_{N/2}(L_{N-1})}{\sqrt{4\pi}}\right]\nonumber\\
&&\qquad\qquad+\frac{\beta}{\sqrt{4\pi}}
\sum_{n=1}^{\infty}\int_{0}^{\infty}t^{-3/2}e^{-(n\beta)^2/4t}
\,\mbox{Tr}\, e^{-tL_{N-1}}\;dt
\label{PFult}\end{eqnarray}
where $\bar\mu=2\ln(e\mu/2)$,
$\bar\zeta$ is the finite part, in the given point,
of the spatial section of the $\zeta$-function $\zeta$
and $K_{N/2}(L_{N-1})$ is the coefficient of $\sqrt{t}$
in the asymptotic expansion
\cite{minakshisundarampleijel48,seeley67,seeley69}
\begin{eqnarray}
\,\mbox{Tr}\, e^{-tL_{N-1}}\sim\sum_{n=0}^{\infty}
K_\frac{n}{2}(L_{N-1})\;t^{\frac{n-(N-1)}{2}}
\label{heat}\end{eqnarray}
Obviously Eq.~(\ref{PFult}) separates into a zero-temperature
contribution, the terms in brackets, and into contributions from the
excited states. As it is well known, only the zero-temperature part needs
renormalization, thus the finite temperature contribution is
independent of the renormalization scale $\mu$.
\section{The role of the conical singularity: vacuum energy}
\label{S:VE}
Here we derive the expressions for the renormalized vacuum
energy for a massive scalar field on
$\mbox{${\rm I\!R }$}^{N-2}\times C_{\gamma}$.
Before doing that, let us first consider in
some detail the heat-kernel and the $\zeta$-function of the relevant
operator on the cone $C_{\gamma}$. The corresponding quantities on the
product manifold $\mbox{${\rm I\!R }$} ^{N-2}\times C_{\gamma}$ are then easily
obtained.

Let us define the manifold $C_{\gamma}$ using global coordinates
$\bar r =(r,\varphi )$ with the range $r\in [0,\infty )$,
$\varphi\in\mbox{${\rm I\!R }$}$, and
with the metric
\begin{eqnarray}
ds^2=dr^2+r^2\,\,d\varphi^2\label{31}
\end{eqnarray}
where we identify $\varphi\sim\varphi+n\gamma$, $n\in\mbox{${\rm I\!N }$}$. The
relevant
operator for a free massive scalar field is
\begin{eqnarray}
L_c=-\Delta _c+m^2\label{32}
\end{eqnarray}
with periodic boundary conditions in the angular variable, where in
the metric (\ref{31}) the Laplace-Beltrami operator is given in the
form
\begin{eqnarray}
\Delta_c=\partial_r^2+\frac 1 r \partial _r+\frac 1 r^2\partial
_{\varphi}^2\label{33}
\end{eqnarray}
A complete set of normalized eigenfunctions with
\begin{eqnarray}
L_c\psi_{n,\nu}(r,\varphi )=(\nu^2+m^2)\psi_{n,\nu}(r,\varphi)\label{34}
\end{eqnarray}
may then be determined to be
\begin{eqnarray}
\psi_{n,\nu}(r,\varphi)=\frac 1 {\sqrt{\gamma}}e^{i\frac{2\pi
n}{\gamma}\varphi}J_{\frac{2\pi|n|}{\gamma}}(r\nu)\label{35}
\end{eqnarray}
with $J_k$ the regular Bessel function.
Using this set of eigenfunctions it is then possible to determine the
heat-kernel $K_t^{\gamma}(\bar r ,\bar r' )$ and the local and global
$\zeta$-function of
the operator $L_c$. Some details
of the calculation are relegated to the appendix A.
There we derive for the heat-kernel
\begin{eqnarray}
K_{t}^{\gamma}(\bar r,\bar r')&=&\frac{e^{-m^2t-(\bar r-\bar r')^2/4t}}{4\pi
t}\left\{
\sum_{\frac{-\pi-(\varphi-\varphi')}{\gamma}<n\leq\frac{\pi-(\varphi-\varphi')}{\gamma}}
\exp\left[-\frac{rr'}{t}\sin^2\frac{n\gamma+\varphi-\varphi'}{2}\right]
\right.\nonumber\\
&&\qquad\qquad\left.-\sin\frac{2\pi^2}{\gamma}
\int_{-\infty}^{+\infty}
\frac{e^{-\frac{rr'}{t}\cosh^2\frac{u\gamma-i(\varphi-\varphi')}{2}}}
{\cosh(2\pi u)-\cos\frac{2\pi^2}{\gamma}}du\right\}
\label{36}
\end{eqnarray}
Considering coincident points $\bar r=\bar r'$,
Eq.~(\ref{36}) simplifies to
\begin{eqnarray}
K_{t}^{\gamma}(\bar r,\bar r)&=&\frac{e^{-m^2t}}{4\pi t}\left\{
\sum_{-\pi <n\gamma\leq\pi}
\exp\left[-\frac{r^2}{t}\sin^2\frac{n\gamma}{2}\right]
\right.\nonumber\\
& &\qquad\qquad \left.
-\int_{-\infty}^{\infty}
\frac{e^{-\frac{r^2}{t}\cosh^2\frac{u}{2}}}
{F(u,\gamma )}\right\}
\label{36a}
\end{eqnarray}
where we introduced
\begin{eqnarray}
F(u,\gamma)=\frac1{\gamma}\frac{\sin\frac{2\pi^2}{\gamma}}
{\cosh\frac{2\pi u}{\gamma}
-\cos\frac{2\pi^2}{\gamma}}\label{38}
\end{eqnarray}
Thus using a well known integral
representation of the Kelvin function $K_{s}$, one finds for the
local $\zeta$-function
\begin{eqnarray}
\zeta_{\gamma}(s,\bar r)
&=&\frac{\Gamma (s-1)}{4\pi\Gamma(s)}m^{2-2s}\label{37}\\
& &+\frac 1 {2\pi \Gamma(s)} {\sum_{-\frac{\pi}{\gamma} <n\leq
\frac{\pi}{\gamma}}}\!\!\!\!^{\prime}
\left|\frac{m}{r\sin\frac{n\gamma}2}
\right|^{1-s}K_{s-1}\left(2r\left|m\sin\frac{n\gamma}2
\right|\right)\nonumber\\
& &-\frac 1 {\pi\Gamma(s)}\int_0^{\infty}du\,\,F(u,\gamma )\left|\frac m
{r\cosh\frac u2}\right|^{1-s}
K_{s-1}\left(2rm\cosh\frac u2\right)\nonumber
\end{eqnarray}
where the prime indicates omission of the summation index $n=0$.
The first term is exactly the Minkowski space contribution.
Normalizing our results to zero for the Minkowski space, we will
neglect this part thus defining $\zeta_{\gamma}^{sing}(s,\bar r )$. For
$\gamma$
rational, the third part is seen to vanish, thus showing clearly the
contributions to $\zeta_{\gamma}^{sing}(s,\bar r )$ resulting from $C_{\gamma}$
not being an orbifold in that case. Furthermore, using the asymptotics
of $K_{s-1}$ one finds that $\zeta_{\gamma}^{sing}(s,\bar r )\sim r^{2s-2}$
for $r\to 0$. This corresponds to the results for a local
$\zeta$-function $\zeta(s|A)$ associated with a second order elliptic
differential operator on a manifold with a boundary, which behaves
like $\delta^{2s-d}$ ($d$ being the spacetime dimension) when the
geodesic distance $\delta$ to the boundary of the manifold
is going to zero. This leads to the well known divergences in local
quantum functions revealed in numerous situations (see for example
\cite{dewitt75,balianduplantier77,balianduplantier78,deutschcandelas79})
and which may be renormalized away by suitable counterterms
\cite{critchleydowkerkennedy80}.

Whereas the local quantities are all well defined, the global
quantities will diverge due to the infinite volume of the cone. This
divergence is equal to the Minkowski space divergence and is given by
the $n=0$ contribution in Eq.~(\ref{36a}). Subtraction of the
Minkowski space contribution leads to the regularized global
$\zeta$-function
\begin{eqnarray}
\zeta_{\gamma}^{sing}(s)=\frac{m^{-2s}}{12}
\left(\frac{2\pi}{\gamma}-\frac{\gamma}{2\pi}\right)\label{39}
\end{eqnarray}
Now everything is prepared to calculate the vacuum energy
for a massive scalar field on $\mbox{${\rm I\!R }$} ^{N-2} \times C_{\gamma}$.
Defining the vacuum energy as the limit $\beta\to\infty$ of
the internal one, Eq.~(\ref{PFult}) directly yields
\cite{cognolavanzozerbini92}
\begin{eqnarray}
E_v&=&\left.-\partial_\beta\ln Z_\beta\right|_{\beta\to\infty}\nonumber\\
&=&\frac{1}{2}\left[\bar\zeta(-1/2|L_{N-1})
-\frac{\bar\mu K_{N/2}(L_{N-1})}{\sqrt{4\pi}}\right]
\label{Ev}\end{eqnarray}
This definition is in agreement with the considerations in
Ref.~\cite{blauvisserwipf88}, where also the relevant remarks concerning
the necessary renormalization may be found.

In order to determine the vacuum energy (\ref{Ev}) we need to
calculate $\bar{\zeta}(-1/2|L_{N-1})$. Using Eq.~(\ref{39}) this
is easily done.
The massless case has to be treated separately, but here we do not
consider it because it was long discussed in
the literature \cite{brownottewill85}.
Calculational details for the massive case are relegated to the
appendix A.

It is convenient to distinguish between the odd and even dimensional
case.
Using Eqs.~(\ref{THK}) and (\ref{ZFren}) we see
that for odd $N\geq3$,
$K_{N/2}(L_{N-1})$ vanishes and
$\zeta_{{\it sing}}(s|L_{N-1})$ is finite for $s=-1/2$, thus
the regularized vacuum energy is simply given by
\begin{eqnarray}
E_{sing}=\frac12\zeta_{sing}(-1/2|L_{N-1})
=-\frac{R_0}{2}
\frac{V_{N-3}\Gamma(1-N/2)}{(4\pi)^{N/2}}
m^{N-2}
\label{VEodd}
\end{eqnarray}
where $V_n$ is the volume of a large box in $\mbox{I$\!$R}^n$,
we introduced
\begin{eqnarray}
R_0=\frac{\pi}{3}
\left(\frac{2\pi}{\gamma}-\frac{\gamma}{2\pi}\right)
\label{}\end{eqnarray}
and
by the suffix $sing$ we indicate the contribution due to the
singularity of those quantities obtained by
subtracting the corresponding quantities on $\mbox{I$\!$R}^N$.
Of course they are vanishing for $\gamma=2\pi$, thus renormalizing the
vacuum energy for the Minkowki spacetime to zero.
As is clearly seen, the sign of the vacuum energy
alternates with the dimension $N$.

In the even case $N\geq 4$, a careful application of
Eq.~(\ref{Ev}) gives
\begin{eqnarray}
E_{sing}=\frac{R_0}{2}
\frac{(-1)^{N/2}V_{N-3}}{(4\pi)^{N/2}\Gamma(N/2)}
m^{N-2}\left( C_{\frac{N}{2}-1}-\ln\frac{m^2}{\mu^2}\right)
\label{VEeven}
\end{eqnarray}
where $C_n=\sum_{k=1}^{n}\frac{1}{k}$.
We note that the vacuum energy vanishes in the massless limit.

\section{The role of conical singularity: finite temperature}
\label{S:FT}

To see how conical singularities modify the
finite temperature theory, we consider as an example
a massive, non interacting scalar field on the manifold
${\cal M}=S^1\times\mbox{I$\!$R}^{N-3}\times{\cal C}_\gamma$
with periodic boundary conditions on the factor $S^1$.
Denoting by $L_{S^1}=-\partial _{\tau}^2 +m^2$
the massive Laplacian on $S^1$,
we have $L_N=L_{S^1}-\triangle_{N-3}-\triangle_c$ and
the heat kernel is the product of the three kernels on $S^1$,
$\mbox{I$\!$R}^{N-3}$ and ${\cal C}_\gamma$. So, using Eq.~(\ref{THK}) we have
\begin{eqnarray}
\,\mbox{Tr}\, e^{-tL_N}=\sum_{n=-\infty}^{\infty}
\frac{V_{N-3}e^{-tM_n^2(\beta)}}{(4\pi t)^{(N-3)/2}}
\left(\frac{V({\cal C}_\gamma)}{4\pi t}+\frac{R_0}{4\pi}\right)
\label{xx}\end{eqnarray}
where $M_n^2(\beta)=m^2+(2\pi n/\beta)^2$ has been put.

Taking the Mellin transform of Eq.~(\ref{xx}) we have
\begin{eqnarray}
\zeta_\beta(s|L_N/\mu^2)=
\frac{\mu^{2s}V_{N-3}}{\Gamma(s)(4\pi)^{\frac{N-1}{2}}}
\left[ V({\cal C}_\gamma)f(s-\frac{N-1}{2})
+R_0f(s-\frac{N-3}{2})\right]
\label{}\end{eqnarray}
The function
\begin{eqnarray}
f(s)=\Gamma(s)\zeta(s|L_{S^1})=\sum_{n=-\infty}^{\infty}\int_0^{\infty}dt\,\,
t^{s-1}\exp\left\{-m^2 t-\left(\frac{2\pi n}{\beta}\right)^2 t\right\}
\label{42a}
\end{eqnarray}
has simple poles at $s=\frac12-k$
with residues $(-1)^k\beta m^{2k}/\sqrt{4\pi}k!$
($k=0,1,\dots$) and it has the asymptotic expansion
for $\beta\to\infty$
$f(s)\sim m\beta\Gamma(s-1/2)/\sqrt{4\pi}m^{2s}$.

For the logarithm of the finite temperature
partition function then we get
\begin{eqnarray}
\ln Z_\beta=\frac{V_{N-3}}{2(4\pi)^{\frac{N-1}{2}}}
\left[ V({\cal C}_\gamma)f(\frac{1-N}{2})
+R_0f(\frac{3-N}{2})
\right]
\label{ZFodd}\end{eqnarray}
valid for odd $N\geq3$, while for even $N\geq4$ it reads
\begin{eqnarray}
\ln Z_\beta&=&\frac{V_{N-3}}{2(4\pi)^{\frac{N-1}{2}}}
\left\{ V({\cal C}_\gamma)\left[\bar f(\frac{1-N}{2})
+\beta m^N(\gamma+\ln\mu^2)
\frac{(-1)^{N/2}}{\sqrt{4\pi}\Gamma(1+N/2)}\right]\right.\nonumber\\
&&\qquad\qquad\left.+R_0\left[\bar f(\frac{3-N}{2})
-\beta m^{N-2}(\gamma+\ln\mu^2)
\frac{(-1)^{N/2}}{\sqrt{4\pi}\Gamma(N/2)}\right]
\right\}
\label{ZFeven}\end{eqnarray}
$\bar f(s)$ being the finite part of $f(s)$ in the given point.
Taking the derivative with respect to $\beta$ and the limit
$\beta\to\infty$ we again recover Eqs.~(\ref{VEodd}) and (\ref{VEeven}).

The results
Eqs.~(\ref{ZFodd}) and (\ref{ZFeven}) are exact and clearly show the
contribution due to the conical singularity vanishing if $\gamma =2\pi$.
Approximations valid for low or high temperature may be obtained by
using well known techniques. We will not do so explicity, but content
ourselves with some remarks.

First of all, performing a Poisson resummation in Eq.~(\ref{42a})
the low temperature expansion of Eqs.~(\ref{ZFodd}) and
(\ref{ZFeven}) is obtained.

For the discussion of the high-temperature limit let us utilize the general
results of \cite{dowkerkennedy78} (see also
\cite{dowkerschofield88,dowkerschofield89,dowkerschofield90,kirsten91b,bytsenkovanzozerbini92}).
There it has been shown, that the relevant quantities for the high
temperature expansion are the heat-kernel coefficients of the operator
(for the considered case) $-\Delta_{N-3}-\Delta_c+m^2$. In Appendix
B we showed, that contributions only due to the existence of the
conical singularity arise, thus altering all orders of the expansion
apart from the leading Planckian term. Thus, in the same way a
boundary changes the high-temperature behaviour of the theory, also a
non-smoothness of a manifold changes this behaviour as exemplified by
the cone.

In addition to finite temperature one may also be interested to
introduce finite densities into the theory thus considering a charged
scalar field. Of course it is once more possible to derive the
equations corresponding to Eqs.~(\ref{ZFodd}) and (\ref{ZFeven}).
However, in that context one is especially interested in the
phenomenon of Bose-Einstein condensation at high temperature, which
may be discussed already using the general formalism given recently by
Toms \cite{toms92,toms93}. Utilizing his results, it may be shown that
the critical temperature at which the gas of a free charged scalar
fields in the given spacetime will condensate is given by
\begin{eqnarray}
T_C=\left(\frac{\pi ^{\frac N 2}Q}{2\zeta_R (N-2)\Gamma\left(\frac N
2\right)m^2}\right)^{\frac 1 {N-2}}\label{bose}
\end{eqnarray}
(with the thermal average of the charge density $Q$), thus not
changing the Minkowski spacetime result derived in
\cite{haberweldon81}. This is essentially due to the fact, that the
smallest eigenvalue of the Laplacian on the cone and in Minkowski
spacetime are equal.

Another important consequence of the additional contributions due
to conical singularities is connected with the conformal anomaly.
In flat Minkowski spacetime the trace of the energy-stress tensor for
a free massless particle is of course vanishing. As is well known
\cite{birelldavies82} for a general spacetime it is proportional to
the coefficient $K_{\frac N 2}$ in Eq.~(\ref{heat}). As we have shown in
Appendix B, $K_1$ depends on the angle $\gamma$ and is
nonvanishing for $\gamma\neq 2\pi$, thus we find an anomalous trace for
the scalar field on the cone. Having in mind, that the angle $\gamma$ in
the presence of horizons may present temperature (see Sec.~\ref{S:RW}),
we explain the temperature dependence of the anomaly recently found in
\cite{fursaevmieleunv}
in a static de Sitter spacetime. As we will explain more detailed in the
Conclusions, this result is to be expected because such spacetimes
may be interpreted to have temperature dependent curvature tensors
\cite{balasinnachbagauerunv} and
consequently also a temperature dependent stress tensor anomaly.

Both Eqs.~(\ref{ZFodd}) and (\ref{ZFeven}) are valid for massive
fields.
In the massless case, disregarding the null eigenvalue
(here it is equivalent to omit the zero temperature contribution),
one has $f(s)=2\Gamma(s)(\beta/2\pi)^{2s}\zeta_R(2s)$,
$\zeta_R(s)$ being the usual
Riemann $\zeta$-function, and so one obtains the more explicit formulae
for the logarithm of the thermodynamic partition function
\begin{eqnarray}
\ln Z_\beta&=&\frac{V_{N-3}}{(4\pi)^{\frac{N-1}{2}}}
\left[\frac{V({\cal C}_\gamma)}{\Gamma(\frac{N+1}2)}
\left(\frac{2\pi}{\beta}\right)^{N-1}\zeta'_R(1-N)\right.\nonumber\\
&&\qquad\qquad\left.+\frac{R_0}{\Gamma(\frac{N-1}2)}
\left(\frac{2\pi}{\beta}\right)^{N-3}\zeta'_R(3-N)
\right]
\label{}\end{eqnarray}
\begin{eqnarray}
\ln Z_\beta&=&\frac{V_{N-3}}{(4\pi)^{\frac{N-1}{2}}}
\left[ V({\cal C}_\gamma)\left(\frac{2\pi}{\beta}\right)^{N-1}
\Gamma(\frac{1-N}2)\zeta_R(1-N)\right.\nonumber\\
&&\qquad\qquad\left.+R_0\left(\frac{2\pi}{\beta}\right)^{N-3}
\Gamma(\frac{3-N}2)\zeta_R(3-N)\right]
\label{}\end{eqnarray}
for odd and even $N$ respectively. Here
comparing with the flat spacetime results \cite{actor86},
the comments of the influence
of the conical singularity
on the high-temperature behaviour are made explicit at subleading
order.

As a simple application of the latter equation, we compute
the free energy for a massless scalar field on the
manifold ${\cal M}=S^1\times\mbox{I$\!$R}\times{\cal C}_\gamma$.
By using definition $F=-\ln Z_\beta/\beta$ we directly get
\begin{eqnarray}
F=-V_1\left[
\frac{V({\cal C}_\gamma)\pi^2}{90\beta^4}
+\frac{R_0}{24\beta^2}\right]
\label{sarah}\end{eqnarray}
where once more we see that the singularity gives a contribution
at subleading order
proportional to $T^2$ to the free energy. A formula very close to the
previous one is valid for the internal energy. In fact one has
\begin{eqnarray}
E=V_1\left[
\frac{V({\cal C}_\gamma)\pi^2}{30\beta^4}
+\frac{R_0}{24\beta^2}\right]
\label{}\end{eqnarray}
showing the same structure as Eq.~(\ref{sarah}).

\section{The Rindler wedge}
\label{S:RW}

Now we consider the space-time region
${\cal R}$, which is causally accessible
to a uniformly accelerated observer with acceleration $a$.
The metric can be taken in the form
\begin{eqnarray}
a^2ds^2=-a^2\xi^2dt^2+d\xi^2+dy^2+dz^2
\label{}
\end{eqnarray}
where $0<\xi<\infty$. Although at each point there are ten Killing
vector fields, only six of them generate a global group of isometries of
${\cal R}$. The set $\Sigma$ defined by $\xi =\xi_{0}$ and $t=0$ is a
space-like submanifold which is left fixed by the one parameter group
of isometries generated by the Killing field
\begin{eqnarray}
K=\sinh t\partial_{\xi}+\left(\frac{1}{\xi_{0}}
-\frac{1}{\xi}\cosh t\right)\partial_t
\label{}
\end{eqnarray}
This is unique up to a scale and choice of $\xi_{0}$. In such a situation
the set of all
future directed null geodesics which intersect $\Sigma$ form a so called
bifurcate Killing horizon. One can easily see this is the limit of vision
for an observer with uniform acceleration $1/\xi_{0}$. This is thus the
surface gravity of the horizon. The Rindler metric has such a bifurcate
horizon and precisely for that reason its Euclidean brother,
with $\tau=it$ compactified on $S^1$, describes the product manifold
${\cal M}={\cal C}_\gamma\times\mbox{I$\!$R}^2$, where now $\gamma=a\beta$ is
proportional to the
inverse temperature.
The theory of a quantum field on this manifold should describe a thermal
state in the Fock space built with the Rindler mode functions, and we should
clarify its relation with the corresponding Minkowski state.
At $\beta=\beta_H=2\pi/a$ we know this is just the Minkowski
vacuum while at $\beta=\infty$ it is the Rindler one
($\beta_H$ represents the Hawking inverse temperature).

The $\zeta$-function for a scalar field on such a manifold is biven by
Eq.~(\ref{ZFren}). Here we need the density, then we consider
Eq.~(\ref{Krr'}) in the coincidence limit and
replace $\gamma$ by $a\beta$ and $r$ by $\xi$.
In the massless case, for the free energy we easily get
\begin{eqnarray}
F_{sing}=-\frac12\zeta'_{sing}(0)=-\int_{{\cal M}}
\frac{1}{1440\xi^4\pi^2}\left(\frac{\beta_{H}^{4}}{\beta^4}+
\frac{10\beta^{2}_{H}}{\beta^2}-11\right)\;d^4x
\label{}
\label{}
\end{eqnarray}
where only the contribution due to the singularity has been written.

Now in a thermal state we expect there is no flux of energy-momentum with
respect to the observers defined by the unique
time-like Killing field ${\cal K}=(1,0,0,0)$. So the
energy-momentum tensor must obey $T_{0i}(\gamma ,\xi)=0$. Moreover
it must be traceless since there are no anomalies in Minkowski space.
Hence, in any coordinate system the stress tensor will be
\begin{eqnarray}
T_{ab}(\beta,\xi)=\frac{1}{1440\xi^4\pi^2}
\left(\frac{\beta_{H}^{4}}{\beta^4}+
\frac{10\beta^{2}_{H}}{\beta^2}-11\right)
\frac{1}{{\cal K}^2}\left(g_{ab}-
\frac{4{\cal K}_{a}{\cal K}_{b}}{{\cal K}^{2}}\right)
\label{}
\end{eqnarray}
where ${\cal K}^{2}={\cal K}_{a}{\cal K}^{a}$.
Of course, at the Hawking temperature $T=a/2\pi$ it vanishes since
the $\beta=\beta_H$ thermal state corresponds to Minkowski vacuum.
One can say there is no conical singularity, in this case.
At $\beta=\infty$ it correspond to the vacuum energy in the Rindler wedge.

\section{Renormalization of $\lambda\phi^4$ on ${\cal
M}=\rm\mbox{I$\!$R}^2\times {\cal C}_\gamma$}
\label{S:RLP4}
As the last point of our paper let us concentrate on self-interacting
$\lambda \phi^4$ theory. To ensure renormalizability we restrict to a
four dimensional manifold, namely ${\cal M}=\mbox{${\rm I\!R }$} ^2\times
C_{\gamma}$.

We will be especially interested in the one-loop effective action of
the theory. Because the effective action is a concept well discussed
in the literature (see for example \cite{buchbinderodintsovshapiro92})
the describtion of its evaluation will be very brief. We will mainly follow
\cite{cognolakirstenvanzo93unv}.

In the functional integral perturbative approach, the effective action
is expanded in powers of $\hbar$ as
\begin{eqnarray}
\Gamma[\hat{\phi} ]=S[\hat{\phi}]+\Gamma^{(1)}+\Gamma '\label{61}
\end{eqnarray}
where $S[\hat{\phi} ]$ is the classical action and the one-loop
contribution $\Gamma^{(1)}$ to the action is given by
\begin{eqnarray}
\Gamma^{(1)}=\frac 1 2 \ln\det\frac A {\mu^2}
=-\frac 1 2 \zeta'(0|A/\mu^2)\label{62}
\end{eqnarray}
with the $\zeta$-function $\zeta(s|A/\mu^2)$ of the operator
$A=-\Delta +M^2$, where we introduced the effective mass $M^2=
m^2+\frac{\lambda} 2 \hat{\phi} ^2$. Furthermore, $\Gamma'$
represents higher loop corrections which we are not going to
discuss.

Assuming constant background field $\hat{\phi}$, the analysis of
Sec.~\ref{S:VE} may be used.
Nearly without any further calculation one finds the representation
\begin{eqnarray}
\zeta(s,\bar r|A)
=\frac{V_2}{4\pi (s-1)}\zeta_{\gamma}(s,\bar r)\label{a61}
\end{eqnarray}
with $\zeta_{\gamma}(s,\bar r)$ given in Eq.~(\ref{37})
with the replacement $m\to M$. Using
Eq.~(\ref{39}) the integrated version of Eq.~(\ref{a61})
reads
\begin{eqnarray}
\zeta(s|A)&=&\frac{V_2V({\cal C}_{\gamma})}{16\pi^2
(s-2)(s-1)}M^{2(2-s)}\label{a62}\\
& &+\frac{V_2 R_0}{16\pi^2 (s-1)}M^{2(1-s)}\nonumber
\end{eqnarray}
where the first term corresponds to the result of Coleman and
Weinberg \cite{colemanweinberg73} and the second term is a correction
due to the singularity of the cone. As is well known by now, all terms
appearing in $\zeta(0|A)\sim K_2(x,x)$ need counterterms in order to
perform the renormalization
\cite{oconnorhu84,cognolakirstenzerbini93,cognolakirstenvanzo93unv,wiesendangerwipfunv}.
So we see already here, that due to
\begin{eqnarray}
\zeta(0|A) =\frac{V_2V({\cal C}_{\gamma})}{32\pi^2}M^4-\frac{V_2
R_0}{16\pi^2}M^2\label{63}
\end{eqnarray}
in addition to the Minkowski space counterterms, additional
counterterms are forced by the conical singularity, which will be
determined in the following. First using Eq.~(\ref{a62}) the
effective action is found to be
\begin{eqnarray}
\Gamma(\hat\phi)&\sim&\int\left[-\frac{\hat\phi\triangle\hat\phi}{2}
+\frac{\lambda\hat{\phi}^4}{24}
+\frac{m^2\hat{\phi}^2}{2}\right] d^4x\nonumber\\
& &+\frac1{64\pi^2}\int_{\cal M}
M^4(\hat\phi)\left(\ln\frac{M^2(\hat\phi)}{\mu^2}-\frac32\right)
d^4x\label{a63}\\
& &-\frac1{64\pi^2}\int_{\cal M}
2R_0M^2(\hat\phi)\left(\ln\frac{M^2(\hat\phi)}{\mu^2}-1\right)
\delta(\bar r)\;d^4x \nonumber
\end{eqnarray}
The last term in Eq.~(\ref{a63}) is due to the presence of the
singularity, $\delta(\bar r)$ being the Dirac delta function on
the cone. The result agrees with the adiabatic expansion in terms of
heat-kernel coefficients presented in \cite{cognolakirstenvanzo93unv}.
This is easily seen by using the heat-kernel coefficients given in
Eq.~(\ref{Krr}). For the example of the cone, the integrated
coefficients $K_l$ are vanishing for $l>2$ and the adiabatic expansion
terminates leading to Eq.~(\ref{a63}).

In order to renormalize all coupling constants,
to the effective Lagrangian density ${\cal L}$ in Eq.~(\ref{a63})
we have to add the counterterms
\begin{eqnarray}
\delta{\cal L}=\delta\Lambda+\frac{\delta\lambda\hat{\phi}^4}{24}
+\frac{\delta m^2\hat{\phi}^2}{2} +\delta \epsilon_0 R
+\frac 1 2 \delta \xi R \hat{\phi} ^2\label{64}
\end{eqnarray}
where we introduced $R=6R_0 \delta (\bar r )$.

Before determining the counterterms explicitly let us give the
motivation for
the chosen scheme. Apart from the conical singularity,
which is normally excluded from the manifold, the
curvature of the cone is obviously vanishing. However,
in a recent article \cite{balasinnachbagauerunv},
the notion of tensors on smooth manifolds
has been generalized
to tensor-distributions on
manifolds with singular points thus finding a curvature distribution
in these singular points, for
example for the cone this distribution is given by
the $R$ introduced above
(from that point of view, we are studying the minimally coupled case).
The very nice feature in regarding $6R_0 \delta
(\bar r )$ as the curvature of the cone is now, that one can take over the
renormalization scheme from smooth manifolds
\cite{colemanweinberg73,oconnorhu84} without changes.
Thus the necessary counterterms are given as in Eq.~(\ref{64}), the last two
terms being concentrated on the tip of the cone. Higher order terms in
the curvature are not needed for the present case
due to the results
presented in Appendix B.

After these remarks,
going on as usual, one imposes
the renormalization conditions
\cite{colemanweinberg73,oconnorhu84}
\begin{eqnarray}
0 &=&{\cal L}\left|_{\hat{\phi} =\varphi_0,R_0=0}\right.\nonumber\\
\lambda &=&\frac{\partial^4{\cal L}}{\partial\hat{\phi}^4}
            \left|_{\hat{\phi}=\varphi_1,R_0=0}\right.\nonumber\\
m^2 &=&\frac{\partial^2{\cal L}}{\partial\hat{\phi}^2}
            \left|_{\hat{\phi}=0,R_0=0}\right.\label{65}\\
0&=&\frac{\partial^3{\cal L}}{\partial\hat{\phi}^2\partial R}
            \left|_{\hat{\phi}=\varphi_2,R_0=0}\right.\nonumber\\
\epsilon &=&\frac{\partial{\cal L}}{\partial R}
            \left|_{\hat{\phi}=0,R_0=0}\right.\nonumber
\end{eqnarray}
and explicitly gets the counterterms
\begin{eqnarray}
64\pi^2\delta\Lambda &=& -64\pi^2\left(\frac{m^2\varphi_0^2}{2}
     +\frac{\lambda\varphi_0^4}{24}\right)
     +\frac{\lambda^2\varphi_0^4}{3M_1^4}(M_1^2-m^2)(2M_1^2+m^2)
      \nonumber\\
   &&+M_0^4\left(\ln\frac{M_1^2}{M_0^2}+\frac32\right)
     -2m^2M_0^2\left(\ln\frac{M_1^2}{m^2}+1\right)\nonumber\\
   &&+m^4\left(\ln\frac{M_1^2}{m^2}-\ln\frac{m^2}{\mu^2}+2\right)\nonumber\\
64\pi^2\delta\lambda &=&\lambda^2\left(\frac{8m^4}{M_1^4}
     +\frac{8m^2}{M_1^2}-16
     -6\ln\frac{M_1^2}{\mu^2}\right)\nonumber\\
64\pi^2 \delta m^2
    &=& -2\lambda m^2\left(\ln\frac{m^2}{\mu^2}-1\right)
   \label{66}\\
64\pi^2 \delta \epsilon_0&=&2m^2\left(\xi -\frac 1
6\right)\left[1-\ln\frac{m^2}{\mu^2}\right]\nonumber\\
64\pi^2 \delta \xi &=&-2\lambda\left(\xi-\frac 1
6\right)\left[\ln\frac{M_2^2}{\mu^2}+\frac{\lambda
\varphi_2^2}{M_2^2}\right]\nonumber
\end{eqnarray}
where $M_i^2=m^2+\frac{\lambda}2\varphi_i^2$ has been put.
For the sake of generality, we choosed different values
$\varphi_i$ for the definition of the physical coupling constants.
This is due to the fact that in general they are measured at different
scales, the behaviour with respect to a change of scale being
determined by the renormalization group equations.
In particular, $\varphi_0$ is the true minimum of the potential
and classically it is equal to zero.

Let us stress once more, that due to the conical singularity additional
counterterms concentrated on the tip of the cone are necessary and
have to be included in the classical Lagrangian. This is a direct
consequence of the heat-kernel expansion presented in appendix B. As
is seen, the heat-kernel coefficients contain contributions
concentrated in the singularity $r=0$ of the cone. Now, the one-loop
renormalization counterterms are completely determined by the
coefficient $K_2(x,x)$
\cite{cognolakirstenvanzo93unv,wiesendangerwipfunv},
thus leading to counterterms of the mentioned type.
As we have shortly described, this terms may be seen as resulting from
the curvature of the cone concentrated in the singular point.

After some tedious calculations one finds the renormalized
one-loop contribution to the
effective Lagrangian density in the form
\begin{eqnarray}
64\pi^2{\cal L}_r^{(1)}&=&-32\pi^2m^2\varphi_0^2-\frac{8\pi^2\lambda
\varphi_0^4} 3 +\lambda m^2\varphi_0^2\left(\ln\frac{m^2}{M_0^2}
+\frac 1 2\right)\nonumber\\
& &+m^4\ln\frac{M^2}{M_0^2}-2m^2R_0\delta (\bar r
)\ln\frac{M^2}{m^2}\label{67}\\
& &-\frac{\lambda^2\varphi_0^4} 4 \left[\ln\frac{M_0^2}{M_1^2}-\frac 3
2 -\frac{4(M_1^2 -m^2)(2M_1^2+m^2)}{3M_1^4}\right]\nonumber\\
& &+\left\{R_0\delta (\bar r
)\left[\ln\frac{M_3^2}{M^2}+1+\frac{\lambda\varphi_2^2}
{M_2^2}\right]+m^2\left[\ln\frac{M^2}{m^2}-\frac 1 2\right]\right\}
\lambda \hat{\phi}^2\nonumber\\
& &+\left\{\ln\frac{M^2}{M_1^2}-\frac{25} 6
+\frac{4m^2(m^2+M_1^2)}{3M_1^4}\right\}\frac{\lambda^2\hat{\phi}^4}
4\nonumber
\end{eqnarray}
Eq.~(\ref{67}) clearly shows the influence of the singular point.
Of course, in the limit in which $R_0$
(effectively the curvature) and $m$ go to 0, we obtain the
well known Coleman-Weinberg result \cite{colemanweinberg73}.

\section{Conclusion}
\label{S:C}
In this paper we considered different aspects of the quantum field
theory of a scalar field in the presence of a conical singularity.
Especially we calculated the vacuum energy and the free
energy of a non-interacting scalar field and the effective potential
in a self-interacting $\phi^4$-theory. The presented differences to
the corresponding Minkowski spacetime results may cleary be
led back to the presence of the conical singularity. As was
exemplified especially clear in Sec.~\ref{S:RLP4},
in general the influence of
the singularity may be well understood by attaching non-vanishing
curvature tensors to the cone $C_{\gamma}$ as proposed in
\cite{balasinnachbagauerunv}. Then as a natural consequence these
curvature terms lead to additional contributions in the physical
quantities we calculated.

Realizing that the curvature tensors depend on the angle $\gamma$ which,
in spacetimes with horizons like the Rindler wedge, represents
temperature, one is naturally led to the temperature dependence of the
stress-energy tensor anomaly in the presence of horizons
\cite{fursaevmieleunv}, which is, however, not in conflict with new
results presented for example in \cite{boschifilhonatividade92}.

\acknowledgments {K.~Kirsten is grateful to Prof.~M.~Toller, Prof.~R.~Ferrari
and Prof.~S.~Stringari for the kind hospitality in the Theoretical Group
of the Department of Physics of the University of Trento.}

\appendix{Heat kernel on the cone}
\label{A:HK}

In this appendix we briefly derive two representations for
the heat kernel of the Laplace operator $-\triangle_c$ on the cone,
for a detailed discussion see ref.~\cite{cheeger83}.

The problem was already studied a long time ago by Sommerfeld
\cite{sommerfeld97} and later by Carslaw \cite{carslaw98}
in connection with heat diffusion
and more recently by Dowker \cite{dowker77}
and Brown and Ottewill \cite{brownottewill85}
in connection with quantum field theory in Rindler space
(for an exaustive list of references on the subject we refer the
reader to Ref.~\cite{dowker77}).

Let ${\cal M}_\infty$ be the infinitely sheeted Riemann surface
with global coordinates $\bar r\equiv(r,\varphi)$ running inside
$[0,\infty)\times (-\infty,+\infty)$
and metric $ds^2=dr^2+r^2d\varphi^2$.
Let us further consider the 2-dimensional manifold, ${\cal C}_\gamma$,
which results from the identification $\varphi\sim\varphi+n\gamma$.
If $\gamma=2\pi/j$ ($j\geq1$) it is a true
orbifold with fundamental domain the infinite wedge
$0<r<\infty$, $0<\varphi<\gamma$, otherwise it is not,
but the metric still describe a cone with semi-angle
$\sin^{-1}(\gamma/2\pi)$.
We are interested in the two basic objects for quantum
field theory, namely the heat-kernel and the $\zeta$-function
of the massive Laplace operator on the cone,
$L_c=-\triangle_c+m^2$, and we choose periodic boundary conditions.
To simplify the formulae we set $m=0$.
The presence of mass simply gives rise to
a trivial exponential factor
which we shall take into account when necessary.

Let us study first the heat kernel for $\gamma=\infty$.

A complete set of normalized eigenfunctions is easily found to be
\begin{equation}
\psi= \frac{1}{(2\pi)^{1/2}}e^{ik\varphi}J_{k}(\nu r)
\label{}
\end{equation}
together with its complex conjugate.
Here $k\geq0$ and $\nu^2\geq0$ is the eigenvalue corresponding to $\psi$
and $\psi^*$, while $J_{k}$ is the regular Bessel function.
These states have a density measure
$d\mu=\nu^{-1}\delta(\nu-\nu')$,
so that the heat kernel reads
\begin{equation}
K_{t}^{\infty}(\bar r,\bar r'|-\triangle_c)=\int_{0}^{\infty}
\frac{dk}{\pi}e^{ik(\varphi-\varphi')}
\int_{0}^{\infty}e^{-\nu^2t} J_{k}(\nu r)J_{k}(\nu r')\nu d\nu
\label{}
\end{equation}
This may from a known integral representation of the function $I_{k}(r)$
\cite{gradshteynryzhik65}
as
\begin{equation}
K_{t}^{\infty}(\bar r,\bar r'|-\triangle_c)=\frac{1}{2\pi t}
e^{-\frac{r^{2}+r'^{2}}{4t}}
\int_{0}^{\infty}e^{ik(\varphi-\varphi')}I_{k}\left(\frac{rr'}{2t}\right) dk
\label{}
\end{equation}

The kernel at finite $\gamma$ can now be
obtained either using the periodicity sum over the frequencies
$2\pi n/\gamma$ or using the eigenfunction expansion.
The result is
\begin{eqnarray}
K_{t}^{\gamma}(\bar r,\bar r'|-\triangle_c)=\frac{1}{2\gamma t}
e^{-\frac{r^2+r'^2}{4t}}
\sum_{n=-\infty}^{+\infty}e^{i\frac{2\pi n}{\gamma}(\varphi-\varphi')}
I_{\frac{2\pi|n|}{\gamma}}\left(\frac{rr'}{2t}\right)
\label{Krr'I}
\end{eqnarray}
The latter expression for the heat kernel of the Laplace operator on
the cone, which we obtain as the natural expansion in terms of
eigenfunctions, can be written in more convenient forms by using known
integral representations for $I_\nu$ \cite{gradshteynryzhik65}.
In particular here we use two representations. The first one clearly
isolate the effects related to the (possible) irrationality of $\gamma$,
which causes our cone to be not a good quotient space
(indeed, from this point of view Scott \cite{scott83} describes
the irrational-$\gamma$ cone as an extremely nasty space),
while the second is very convenient from the computational point of
view.

The representations read
\begin{eqnarray}
I_{\nu}(z)=\frac{1}{2\pi}\int_{-\pi}^{\pi}e^{z\cos \theta}\cos \nu \theta
d\theta
-\frac{\sin \nu \pi}{\pi}\int_{0}^{\infty}e^{-z\cosh x-\nu x}dx
\label{repI1}
\end{eqnarray}
valid for $\,\mbox{Re}\, z>0$ and $\,\mbox{Re}\,\nu\geq0$, and
\begin{eqnarray}
I_{\nu}(z)=\frac{e^{-i\pi\nu/2}}{2\pi}
\int_{\Gamma}e^{is\nu+z\sin s} ds
\label{repI2}
\end{eqnarray}
valid for $\,\mbox{Re}\, z>0$ and $-\pi<\mbox{arg}\;z\leq\pi/2$. In the latter
expression, $\Gamma$ is an arbitrary open contour integral in the
complex plane from $\varepsilon-\pi+i\infty$ to $\varepsilon+\pi+i\infty$.

If we plug Eq.~(\ref{repI1}) into Eq.~(\ref{Krr'I}) then we get
the kernel as a sum of two different terms which arise from the
two integrals in Eq.~(\ref{repI1}).
Then using Poisson summation formula we finally obtain
\begin{eqnarray}
K_{t}^{\gamma}(\bar r,\bar r'|-\triangle_c)&=&\frac{e^{-(\bar r-\bar
r')^2/4t}}{4\pi t}\left\{
\sum_{\frac{-\pi-(\varphi-\varphi')}{\gamma}<n\leq\frac{\pi-(\varphi-\varphi')}{\gamma}}
\exp\left[-\frac{rr'}{t}\sin^2\frac{n\gamma+\varphi-\varphi'}{2}\right]
\right.\nonumber\\
&&\qquad\qquad\left.-\sin\frac{2\pi^2}{\gamma}
\int_{-\infty}^{+\infty}
\frac{e^{-\frac{rr'}{t}\cosh^2\frac{u\gamma-i(\varphi-\varphi')}{2}}}
{\cosh(2\pi u)-\cos\frac{2\pi^2}{\gamma}}du\right\}
\label{Krr'1}
\end{eqnarray}
where the maximum value of the $n$ summation index is equal to
$\frac{\pi-(\varphi-\varphi')}{\gamma}$ when this quantity is an integer.

{}From Eq.~(\ref{repI2}), after some calculations, we get the
known integral represetation in the complex plane \cite{carslaw19}
\begin{eqnarray}
K_t^\gamma(\bar r,\bar r'|-\triangle_c)=\frac{e^{-(\bar r-\bar r')^2/4t}}{4\pi
t}
\left[ 1+\frac1\gamma\int_\Gamma '
\frac{e^{-(rr'/t)\sin^2(z/2)}}
{1-e^{-\frac{2\pi i}{\gamma}(z+\varphi'-\varphi)}}dz\right]
\label{Krr'}
\end{eqnarray}
Here $\Gamma '$ is an arbitrary contour integral composed of
two-branches, the first one from $\pi+i\infty$ to
$\pi-i\infty$, intersecting the positive real axis very close to
the origin and the second being the specular copy of the previous one.
For later convenience, in Eq.~(\ref{Krr'})
we have emphasized the contribution due to the conical singularity.

By integrating the latter representation on the cone,
we easily obtain the trace of the kernel in the form
\begin{eqnarray}
\,\mbox{Tr}\, e^{-t\triangle_c}=\frac{V({\cal C}_\gamma)}{4\pi t}
+\frac{1}{12}\left(\frac{2\pi}{\gamma}-\frac{\gamma}{2\pi}\right)
\label{THK}
\end{eqnarray}
where $V({\cal C}_\gamma)$ is the (infinite) volume of the cone.
Of course, only  Weyl \cite{chavel84} and Kac \cite{kac66} contributions
are present.

If we consider the more general case of a massive theory on the manifold
${\cal M}=\mbox{I$\!$R}^{N-2}\times{\cal C}_\gamma$, then the previous
expression has to
be multiplied by the factor $(4\pi t)^{-(N-2)}V_{N-2}\exp(-tm^2)$,
the exact heat-kernel of $-\Delta +m^2$ in $\mbox{${\rm I\!R }$} ^{N-2}$.
So the contribution to the $\zeta$-function due to the singularity,
that means after subtraction of the Minkowskispace part,
can be computed by a Mellin transform, the result being
\begin{eqnarray}
\zeta_{sing}(s|L_N)
=\frac{\pi}{3}\left(\frac{2\pi}{\gamma}-\frac{\gamma}{2\pi}\right)
\frac{V_{N-2}\Gamma(s-\frac{N-2}{2})}{(4\pi)^{N/2}\Gamma(s)}
m^{-2(s-\frac{N-2}{2})}
\label{ZFren}
\end{eqnarray}

\appendix{Singular heat kernel expansion}
\label{A:HKE}

Here we derive the asymptotic expansion for the heat kernel
derived in the previous Appendix \ref{A:HK}.
Of course we expect such an expansion to be a
distribution concentrated on the tip of the cone (apart from the first
term which corresponds to the kernel on $\mbox{I$\!$R}^2$).
For the kernel we use the second integral representation,
Eq.~(\ref{Krr'}).

In order to obtain the parametrix for the trace,
we expand $\exp(-\sigma r^2)$ in terms of distributions on the
cone.
We consider test functions $\phi(r,\varphi)$ with $\gamma$ periodicity
with respect to $\varphi$, and define \cite{gelfandvilenkin64}
\begin{eqnarray}
\bar\phi(r)=\frac1\gamma\int_0^\gamma\phi(r,\varphi)d\varphi
\end{eqnarray}

The Taylor expansion for $\bar\phi(r)$ is given by
the Pizzetti formula \cite{gelfandvilenkin64}
\begin{eqnarray}
\bar\phi(r)=\sum_k\frac{\triangle_c^k\phi(0,0)}{(2^kk!)^2}
\end{eqnarray}
Using the latter relation we easily get
\begin{eqnarray}
\frac{\alpha\sigma}{\pi}e^{-\sigma r^2}
=\sum_k\frac{\triangle_c^k\delta(\bar r)}{k!(2\sigma)^k}
=\exp(\frac{\triangle_c}{2\sigma})\;\delta(\bar r)
\end{eqnarray}
where $\alpha=2\pi/\gamma$ has been set.
If we use this expression in Eq.~(\ref{Krr'}) for
$\bar r=\bar r'$ we can make the complex integral by the residue method.
The final result for the asymptotic behaviour for $t\to 0$ of the
heat-kernel reads
\begin{eqnarray}
K_t^\gamma(\bar r,\bar r|-\triangle_c)=\frac{1}{4\pi t}\left[ 1
+\sum_{k=0}^\infty\frac{R_k\triangle_c^k\delta(\bar r)}{2^kk!}\;t^{k+1}\right]
\label{Krr}
\end{eqnarray}
where $R_k$ is the residue in zero of the function
$-i\pi[\sin(z/2)]^{-2(k+1)}(1-e^{-i\alpha z})^{-1}$.
In particular we have
\begin{eqnarray}
R_0=\frac{\pi(\alpha^2-1)}{3\alpha};\qquad\qquad
R_1=\frac{\alpha^2+11}{15}R_0;\qquad\qquad
R_2=\frac{2\alpha^4+23\alpha^2+191}{315}R_0;
\label{}\end{eqnarray}
It is seen that by integrating (\ref{Krr}) in order to obtain the
integrated heat-kernel coefficient, only the term $k=0$ survives. Thus
in the calculated physical quantities only $R_0$ is present.

Now we want to extend the expansion in Eq.~(\ref{Krr}) to a more
general operator $L_4=-\triangle+m^2+X(x)$, on ${\cal
M}=\mbox{I$\!$R}^2\times{\cal C}_\gamma$.
To this aim we make the ansatz
\begin{eqnarray}
K_t^\gamma(x,x'|L_4)\sim K_t^\gamma(x,x'|-\triangle)\:
e^{-tM^2}\sum_n b_n(x,x')\; t^n
\label{Kxx'}
\end{eqnarray}
where $M^2=m^2+X$ is a positve scalar function and
$K_t^\gamma(x,x'|-\triangle)$ is the kernel of the Laplacian
on ${\cal M}$, that is the product of the two kernels
$K_{\mbox{I$\!$R}^2}\:K_{{\cal C}_\gamma}$.
It has to be noted that the diagonal part
$K_t^\gamma(x,x|-\triangle)$ of this kernel differs from
the corresponding one on $\mbox{I$\!$R}^4$ for the addition of an
exponentially vanishing function as $t\to 0$.
This is also true for the derivatives of any orders
with respect to $x$.
This means that the coefficients $b_n$, in the coincidence limit
$x'\to x$, are not modified in the presence of the singularity
(the whole contribution of that being in
$K_t^\gamma(\bar r,\bar r|-\triangle_c)$).
Then we have
\begin{eqnarray}
b_0(x,x)=1;&&\qquad\qquad
b_2(x,x)=-\frac{\triangle X}{6};\nonumber\\
b_1(x,x)=0;&&\qquad\qquad
b_3(x,x)=\frac{(\nabla X)^2}{12}-\frac{\triangle^2 X}{60}
\label{}\end{eqnarray}
which finishes the summary of our results on the heat-kernel expansion
on the cone.

\end{document}